\documentclass{svmult}

\usepackage{bm}
\usepackage{cite}
\usepackage{courier}            
\usepackage[bottom]{footmisc}   
\usepackage{graphicx}           
\usepackage{helvet}             
\usepackage{mathptmx}           
\usepackage{multicol}           

\newcommand{\dslash}{\not{\hbox{\kern-2pt $\partial$}}}

\begin{document}

\title*{
\leftline{{\small {\rm HU-EP-09/57}}}\vspace*{-0.3cm}
\leftline{{\small {\rm LMU-ASC 49/09}}} 
\vspace{3mm} 
Topological Structure of the QCD Vacuum Revealed by Overlap Fermions}
\titlerunning{Topological Structure of the QCD Vacuum Revealed by Overlap
Fermions}

\author{Ernst-Michael Ilgenfritz \and Karl Koller \and Yoshiaki Koma \and Gerrit Schierholz \and Volker Weinberg}
\authorrunning{E.-M.~Ilgenfritz, K.~Koller, Y.~Koma, G.~Schierholz, V.~Weinberg}

\institute{Ernst-Michael Ilgenfritz
\at Institut f\"ur Physik, Humboldt-Universit\"at zu Berlin, 12489 Berlin, Germany \\ \email{ilgenfri@physik.hu-berlin.de}
\and
Karl Koller 
\at Fakult\"at f\"ur Physik, Ludwig-Maximilians-Universit\"at M\"unchen, 80333 M\"unchen, Germany \\ \email{Karl.Koller@lrz.uni-muenchen.de}
\and
Yoshiaki Koma
\at Numazu College of Technology, 3600 Ooka, Numazu-shi, Shizuoka 410-8501, Japan \\ \email{koma@numazu-ct.ac.jp}
\and
Gerrit Schierholz
\at Institut f\"ur Theoretische Physik, Universit\"at Regensburg, 93040 Regensburg, Germany \\ Deutsches Elektronen-Synchrotron DESY, 22603 Hamburg, Germany \\ \email{gsch@mail.desy.de}
\and
Volker Weinberg
\at Leibniz-Rechenzentrum der Bayerischen Akademie der Wissenschaften, \\
85748 Garching b. M\"unchen, Germany \\ \email{Volker.Weinberg@lrz.de}}


\maketitle

\vspace{-1.0cm}
\abstract{Overlap fermions preserve a remnant of chiral symmetry 
on the lattice. They are a powerful tool to investigate the topological 
structure of the vacuum of Yang-Mills theory and full QCD.
Recent results concerning the localization of topological charge and the 
localization and local chirality of the overlap eigenmodes are reported. 
The charge distribution is radically different, if a spectral cut-off for 
the Dirac eigenmodes is applied. The density $q(x)$ is changing from the 
scale-$a$ charge density (with full lattice resolution) to the ultraviolet 
filtered charge density. The scale-$a$ density, computed on the Linux cluster
of LRZ,  has a singular, sign-coherent 
global structure of co-dimension 1 first described by the Kentucky group. 
We stress, however, the cluster properties of the UV filtered topological 
density resembling the instanton picture.
The spectral cut-off can be mapped to a bosonic smearing procedure. 
The UV filtered field strength reveals a high degree of (anti)selfduality
at ``hot spots'' of the action. The fermionic eigenmodes show a high degree 
of local chirality. The lowest modes are seen to be localized in low-dimensional
space-time regions.}  

\section{Introduction: Overlap Fermions and Topological Charge}

\vspace{-0.2cm}
Quantum chromodynamics (QCD) is the theory of strong interactions.
It is formulated in terms of quarks and gluons.
The task is twofold: to describe the composite structure and the high-energy 
interactions of strongly interacting {\it hadrons}, in both cases taking the 
substructure in terms of quarks and gluons into account. One problem for 
theorists and experimentalists is that quarks are permanently {\it confined} 
inside hadrons. Apart from details, the spectrum and symmetries of hadrons are 
dictated by an approximate {\it chiral symmetry} and its spontaneous breaking 
by the interaction via gluons. These properties are unique for QCD as part 
of the standard model and result from the vacuum fluctuations of gluons. 
At a temperature 
of about $160 {\rm~MeV}$ the hadronic world experiences a transition to a 
quark-gluon plasma phase with rather unusual properties. Then, the vacuum 
structure has changed.

Simulations on a space-time lattice are the only {\it ab initio} approach
to these phenomena. This approach has the virtue that also structural 
information on the vacuum fluctuations is accessible. 
On the other hand, there are models 
attempting to give a qualitative understanding. Some of them, like the instanton 
model, were partially successful. The instanton model became challenged more 
recently. A study of how the {\it topological charge density} is distributed in 
space-time and how the gluon field localizes the quarks (in analogy to the 
Anderson effect) is crucial to understand the microscopic mechanisms.

Overlap fermions~\cite{Neuberger:1997fp,Neuberger:1998wv}
possess an exact chiral symmetry on the lattice~\cite{Luscher:1998pq}
and realize the Atiyah-Singer index theorem at a finite
lattice spacing $a$~\cite{Hasenfratz:1998ri}. This is possible because they 
allow a clear distinction between chiral zero modes and non-chiral non-zero modes.
Depending on their chirality, counting of $n_{+}$ or $n_{-}$ zero modes 
determines the topological charge of the gauge field as $Q=n_{-}-n_{+}$.
Furthermore, they give rise to a {\it local} definition of the topological 
charge density~\cite{Niedermayer:1998bi}. Altogether, this makes overlap 
fermions an attractive tool for investigating the chiral and topological 
structure of the QCD vacuum. 

The topological structure attracts attention 
because of the old hope that local excitations contributing to the winding 
number might not only realize the breaking of $\rm U_A(1)$ symmetry but  
simultaneously provide a mechanism for confinement and chiral symmetry breaking. 
The instanton liquid model (ILM) does not fulfill this expectation: it can account
for chiral symmetry breaking but fails to give an explanation of confinement.

When the QCDSF collaboration had started to analyze ensembles of lattice 
configurations with overlap fermions for quenched 
QCD~\cite{Galletly:2003vf,Galletly:2006hq}
and later for QCD with dynamical quarks~\cite{Gockeler:2006vi} by diagonalizing 
the massless overlap Dirac operator and to store an -- until then unprecedent -- 
number of eigenmodes (${\cal O}(150)$ per configuration), the way was open for a 
serious investigation of the structure of 
topological charge~\cite{Koma:2005sw,Ilgenfritz:2005hh,Ilgenfritz:2007xu}. 
Not long before we started this investigation, the instanton model had been 
challenged~\cite{Horvath:2002yn,Horvath:2003yj,Horvath:2005rv} by the 
observation that the topological charge, 
rather than appearing in $4d$ clusters, possesses a global, sign-coherent 
$3d$ membrane-like structure~\cite{Thacker:2006wr,Thacker:2006ue,Thacker:2008dr}.
We shall discuss here what 
remains from the instanton picture.

The overlap Dirac operator $D$ has to fulfill the 
Ginsparg-Wilson~\cite{Ginsparg:1981bj}
equation,
\begin{equation}
\gamma_5\, D^{-1} + D^{-1}\, \gamma_5 = a\, 2\, R\, \gamma_5 \, ,
\label{eq:GinspargWilson}
\end{equation}
with a local operator $R$. This is what maximally can remain of chiral symmetry 
on the lattice. A possible solution -- for any input Dirac 
operator, i.e. for the Wilson-Dirac operator $D_W$ as well -- is the following
zero mass overlap Dirac operator 
\begin{equation}
D(m=0)=\frac{\rho}{a}\,\left( 1 + \frac{D_W}{\sqrt{D_W^{\dagger}\,D_W}}
\right)=\frac{\rho}{a}\,\left( 1 + {\rm sgn}(D_W) \right) \, ,
\label{eq:OverlapDirac}
\end{equation}
with $D_W = M - \frac{\rho}{a}$ where $M$ is the Wilson hopping term 
and $\frac{\rho}{a}$ a negative mass term to be optimized. 
This operator is diagonalized using a variant of the Arnoldi algorithm.
The operator can be improved by projecting 
the Ginsparg-Wilson circle on the
imaginary axis, $\lambda \to \lambda_{\rm imp}$.
The topological density can be defined (with maximal resolution $a$) as
follows:
\begin{equation}
q(x) = - {\rm tr} \left[ \gamma_5 \left( 1 - \frac{a}{2}\,D(m=0;x,x) \right)\, \right] \, .
\label{eq:TopDensI}
\end{equation}
Using the spectral representation of (\ref{eq:TopDensI}) in terms of 
the eigenmodes $\psi_{\lambda}(x)$ 
with eigenvalue $\lambda$, an UV smoothed form of the density can be defined by
filtering,
\begin{equation}
q_{\lambda_{\rm sm}}(x) = - \sum_{|\lambda| < \lambda_{\rm sm}} 
\left( 1 - \frac{\lambda}{2} \right) 
\,\sum_c \left( \psi_{\lambda}^c(x)\, ,\gamma_5\, \psi_{\lambda}^c(x) \right) \, ,
\label{eq:TopDensII}
\end{equation}
summed over color $c$ and with $\lambda_{\rm sm}$ as an UV cut-off. Our intention was to study 
the topological charge density at {\it all scales}, in particular the spectral 
representation in terms of overlap modes defining $q_{\lambda_{\rm sm}}(x)$. 
Also, we had the opportunity to evaluate the topological charge density without 
filtering (i.e. at the resolution scale $a$) on the Linux cluster of the LRZ in
Munich. 
We call 
this the scale-$a$ topological density $q(x)$ in contrast to $q_{\lambda_{\rm sm}}(x)$ 
defined with UV smoothing. (For details see Ref.~\cite{Ilgenfritz:2007xu}.)
\vspace{-0.6cm}

\section{Topological Density with Different Resolution}

\begin{figure}[t]
    \centering
    \includegraphics[scale=0.65]{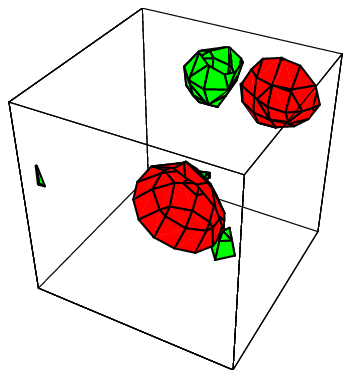}
    \hspace*{0.2cm}
    \includegraphics[scale=0.65]{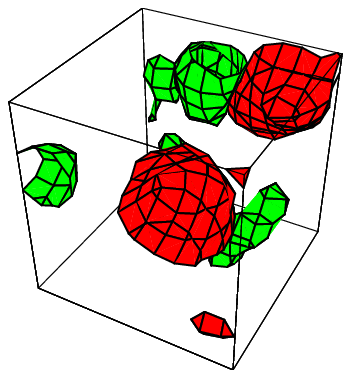}
    \hspace*{0.2cm}
    \includegraphics[scale=0.65]{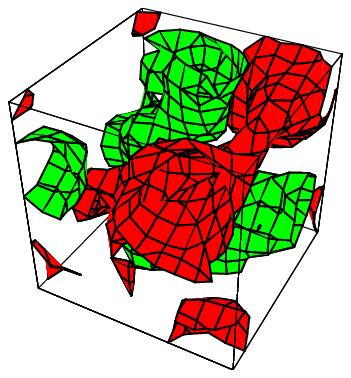}
    \hspace*{0.2cm}
    \includegraphics[scale=0.65]{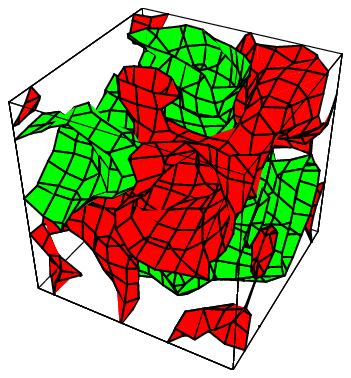}
\caption{Distribution of the topological charge density in a given 
time slice of a $12^3 \times 24$ lattice at $\beta=8.10$:
the spectral cut-off $a\,\lambda_{\rm sm}=$0.14, 0.28, 0.42, 0.56
(from left to right). The isosurfaces are shown for $q(x)=\pm 0.0005$. 
Colors red/green denote the sign of 
topological charge. Fig. from~\cite{Koma:2005sw}. } 
\label{fig:Fig1}
\end{figure}

\vspace{-0.2cm}
Figure~\ref{fig:Fig1} shows the topological density calculated with four 
different smoothing scales $\lambda_{\rm sm}$ given in units of $1/a$. 
The presentation is in the form of isosurfaces at a fixed value of 
$|q(x)|=q_{\rm cut}$. With increasing $\lambda_{\rm sm}$, the number and shape 
of the visible clusters changes. The given configuration has $Q=1$. 
\begin{figure}[ht]
    \centering
    \includegraphics[scale=0.65]{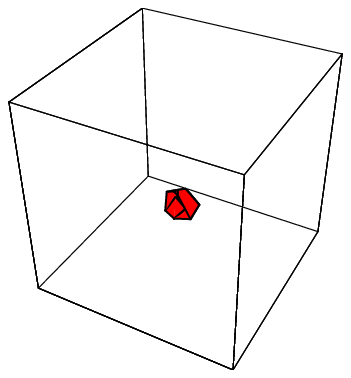}
    \hspace*{1.0cm}
    \includegraphics[scale=0.65]{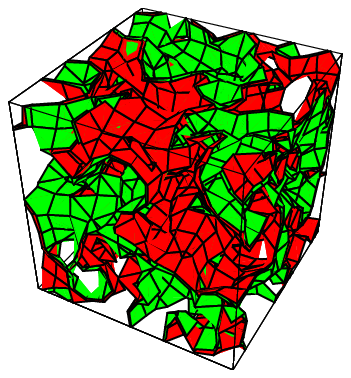}
\caption{The same plot as in Fig.~1 for the zero mode contribution (left) 
and the unfiltered density (right). Fig. from~\cite{Koma:2005sw}. } 
\label{fig:Fig2}
\end{figure}

Therefore exactly one zero mode exists. In Fig.~\ref{fig:Fig2} we show 
two extreme cases, the highly localized contribution of the single zero 
mode alone (left) and the scale-$a$ topological density (right). For the 
same value of $q_{\rm cut}$ the isosurfaces of the latter fill the whole volume, 
separating essentially a single positive from a single negative cluster. 
The net charge is naively expected to reside in an unpaired instanton 
(located at the zero mode).  Actually, however, the positive and negative 
charge (differing by one) is globally distributed over the two clusters. 
\vspace{-0.6cm}

\section{Cluster Analysis}

\vspace{-0.2cm}
More quantitative information can be obtained from a cluster analysis.
The value $q_{\rm cut}$, that has characterized the isosurfaces in Figs. \ref{fig:Fig1} 
and \ref{fig:Fig2}, is now treated as a running parameter. The region characterized 
by $|q(x)| > q_{\rm cut}$ (i.e. the interior of the isosurfaces) consists of some number 
of mutually disconnected clusters. We have classified them according to their size 
and other properties, for example their fractal dimension.
\begin{figure}[t]
    \centering
    \includegraphics[scale=0.43]{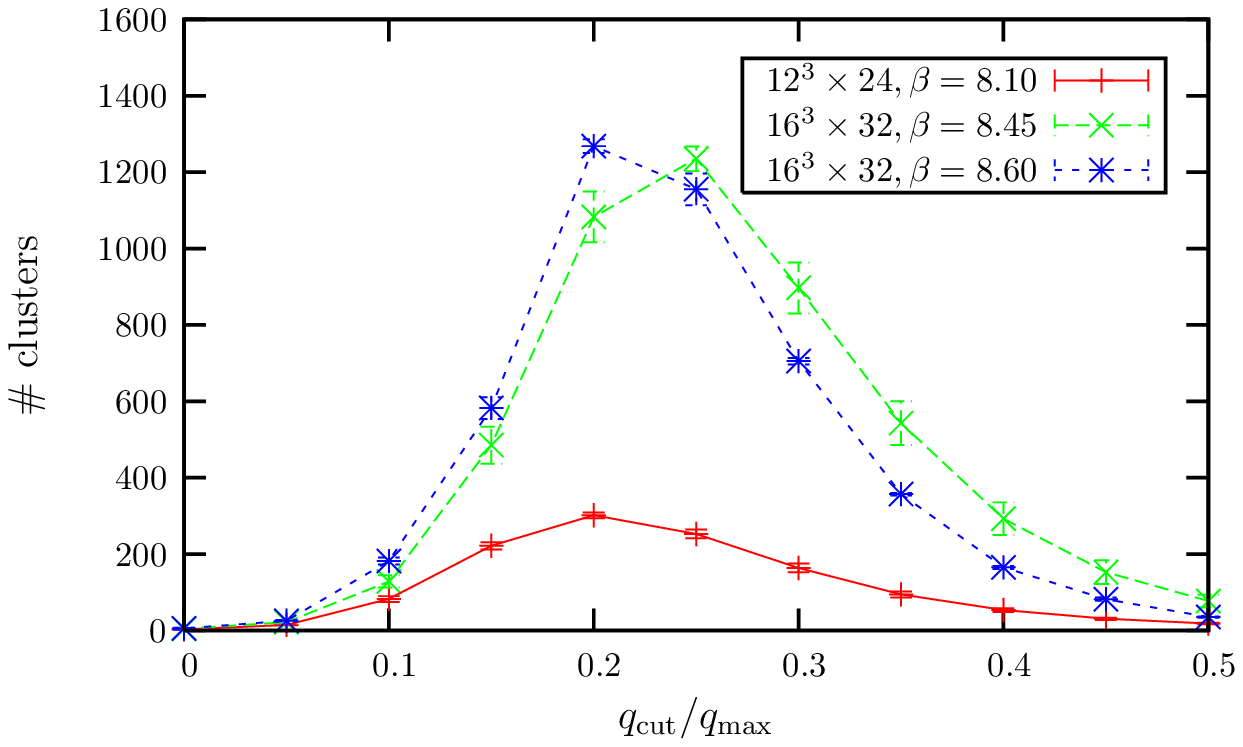}
    \hspace*{0.2cm}
    \includegraphics[scale=0.43]{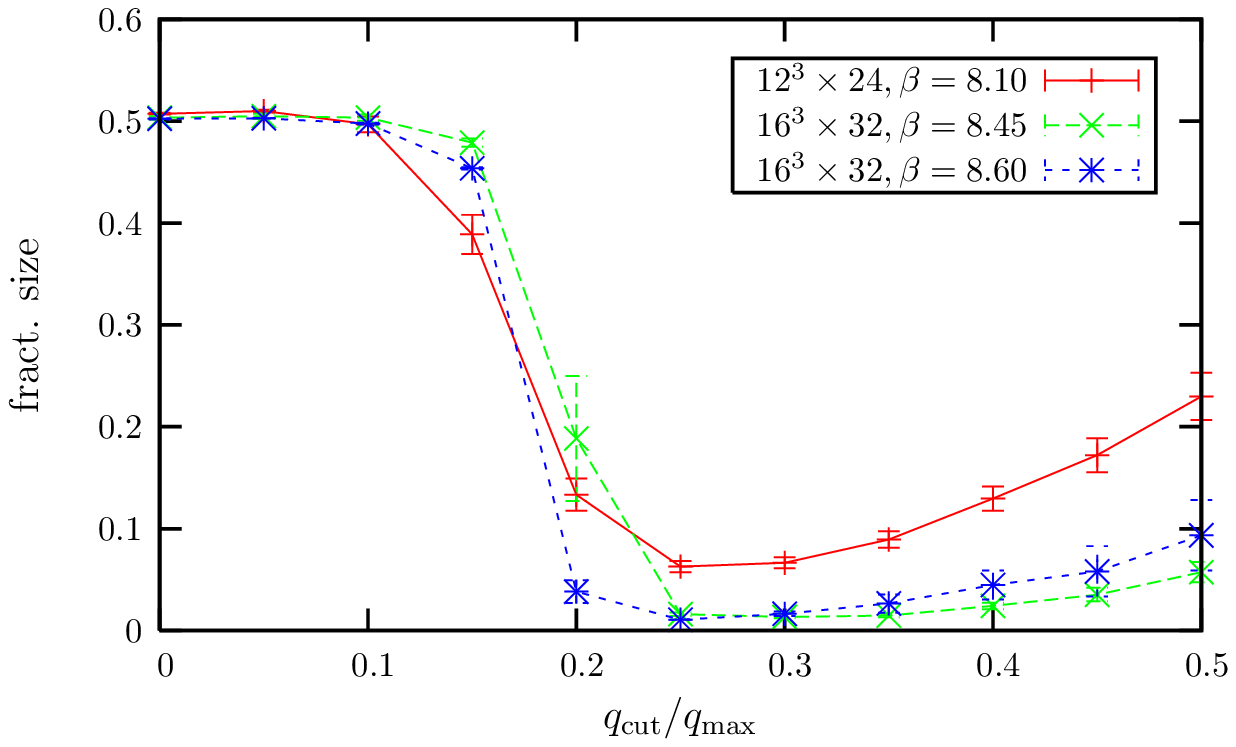}\\
    (a) ~~~~~~~~~~~~~~~~~~~~~~~~~~~~~~~~~~~~~~~ (b)\\
    \includegraphics[scale=0.43]{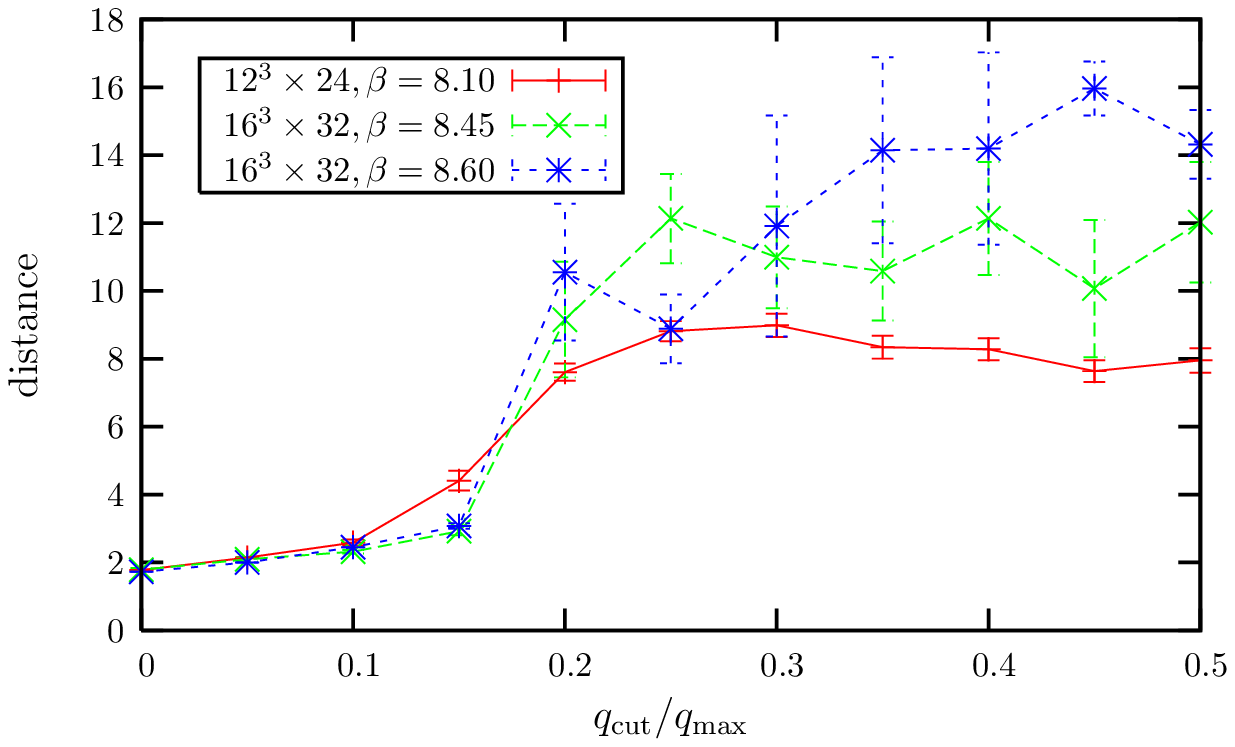}
    \hspace*{0.2cm}
    \includegraphics[scale=0.43]{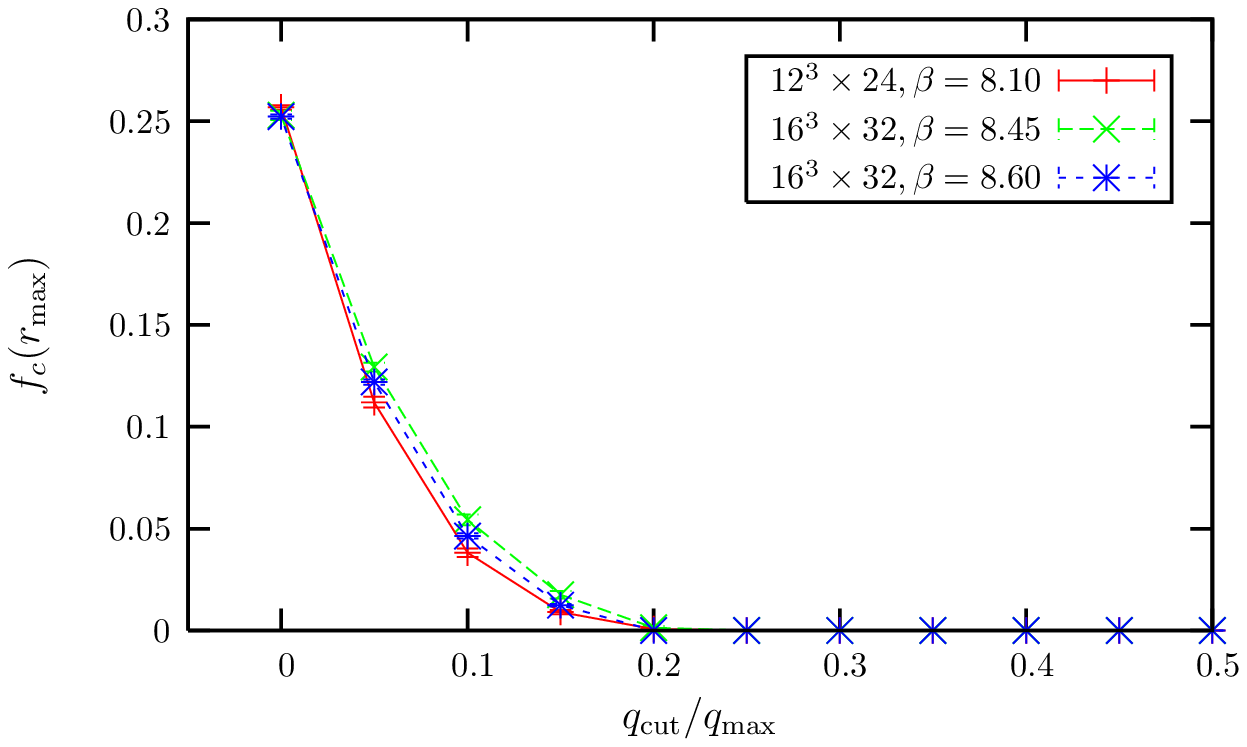}\\ 
    (c) ~~~~~~~~~~~~~~~~~~~~~~~~~~~~~~~~~~~~~~~ (d)
\caption{Cluster analysis of the unfiltered topological density: 
the $q_{\rm cut}$-dependence is shown  
(a) of the number of separate clusters, 
(b) of the volume of the largest cluster relative to all clusters, 
(c) of the distance between the two largest clusters in lattice units and 
(d) of the connectivity describing cluster percolation~\cite{Ilgenfritz:2007xu}.
The data is plotted for the $12^3 \times 24$ lattice configurations at
$\beta=8.10$ (red), for the $16^3 \times 32$ lattice
configurations at $\beta=8.45$ (green) and for the
$16^3 \times 32$ lattice configurations at $\beta=8.60$ (blue).
Fig. from~\cite{Ilgenfritz:2007xu}.}
\label{fig:Fig3}
\end{figure}
Results of the cluster analysis are presented in Fig.~\ref{fig:Fig3}. 
In Fig. 3a we show the number of clusters as function of $q_{\rm cut}/q_{\rm max}$ 
for 
three lattices characterized by different coarseness. Except for the coarsest 
lattice with $\beta=8.1$ ($a=0.142 {\rm~fm}$), the behavior is similar. Around 
$q_{\rm cut}/q_{\rm max}=0.2 \ldots 0.25$ the multiplicity of clusters reaches a 
maximum with a cluster density of ${\cal O}(75 {\rm~fm}^{-4})$ for 
$\beta=8.45$ ($a=0.105 {\rm~fm}$). The density is even higher for the finer 
lattice with $\beta=8.60$ ($a=0.096 {\rm~fm}$). In the limit 
$q_{\rm cut}/q_{\rm max} \to 0$ 
the number of clusters reduces to the two oppositely charged global clusters. 
Figure 3b shows the volume fraction of the biggest cluster relative to all clusters. 
Except for the coarsest ($\beta=8.1$) lattice, above $q_{\rm cut}/q_{\rm max}=0.2$ 
where 
the number of clusters is still growing towards the maximum, this fraction is small 
($ < 10$ \%). It changes rapidly below $q_{\rm cut}/q_{\rm max}=0.2$ and approaches 
$50$ \% for $q_{\rm cut}/q_{\rm max} < 0.1$ and remains so in the limit when only the 
two biggest clusters remain. The distance between the two largest clusters 
is shown in Fig. 3c. The ``distance'' is defined as the {\it maximum} taken over 
all points in one cluster of the {\it minimal distance} to any point of the other 
cluster. It approaches $2\, a$ for $q_{\rm cut}/q_{\rm max} < 0.1$, meaning that the 
two clusters are closely entangling each other. There are no points deep in 
the interior, such that the clusters must be considered to be of lower dimension 
than $4d$. Figure 3d shows the connectivity (explained in~\cite{Ilgenfritz:2007xu}) 
of the biggest 
cluster, describing the ability to span the whole lattice. The connectivity signals 
the onset of percolation which begins below $q_{\rm cut}/q_{\rm max}=0.2$, too. All 
this confirms the description of ``melting instantons'' given earlier by 
the Kentucky group~\cite{Horvath:2002yn,Horvath:2003yj,Horvath:2005rv}.

\begin{figure}[t]
    \centering
    \includegraphics[scale=0.22]{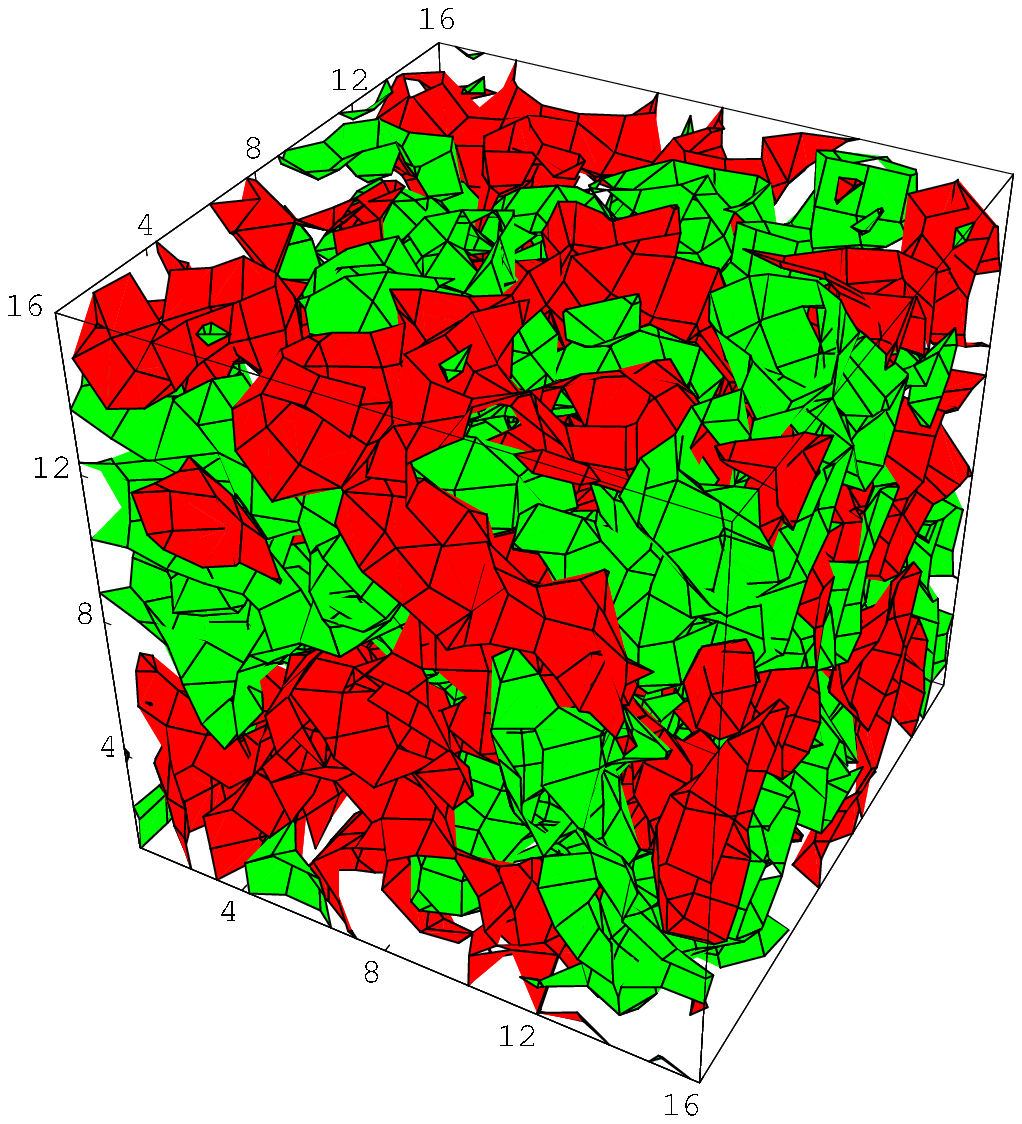}
    \hspace*{0.2cm}
    \includegraphics[scale=0.22]{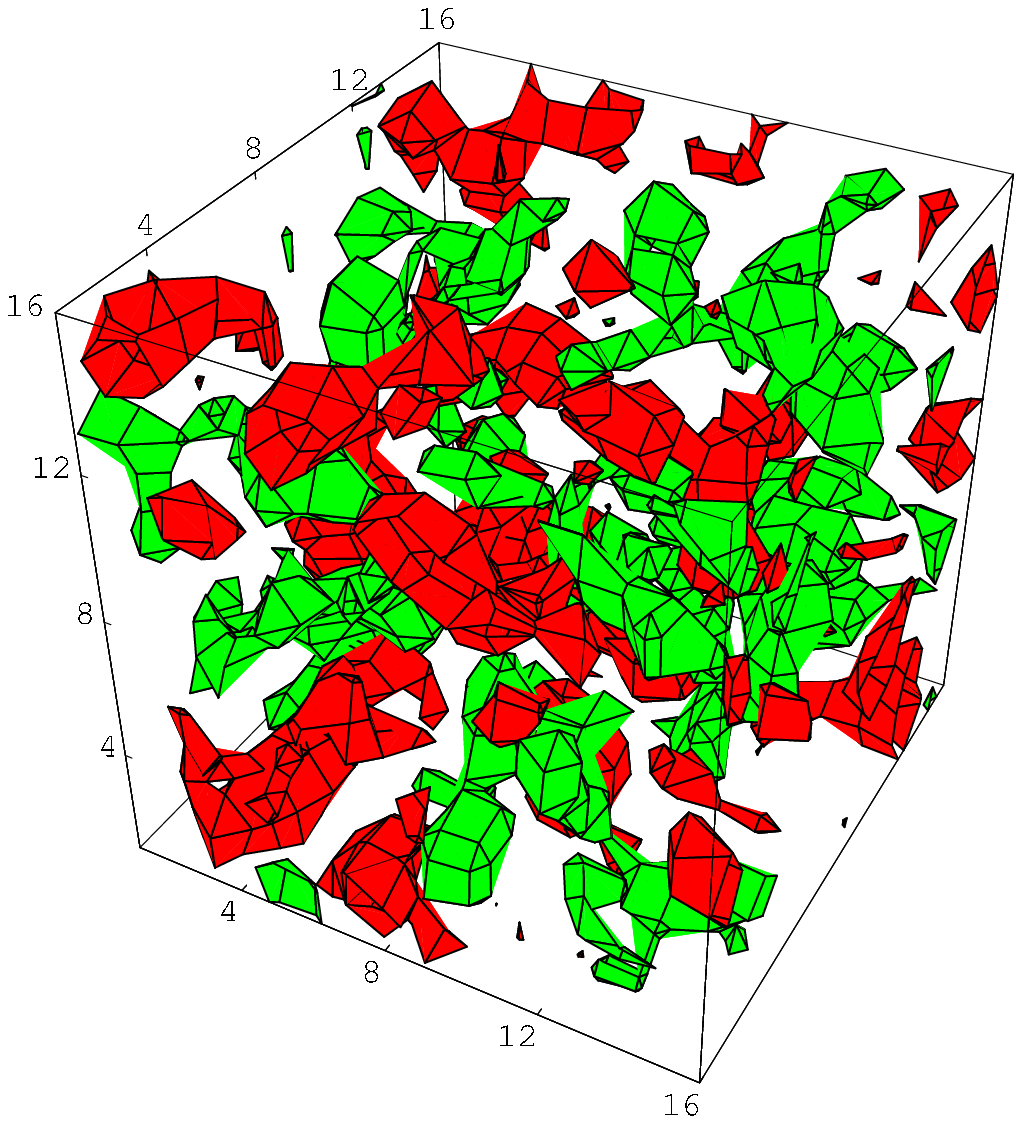}
    \hspace*{0.2cm}
    \includegraphics[scale=0.22]{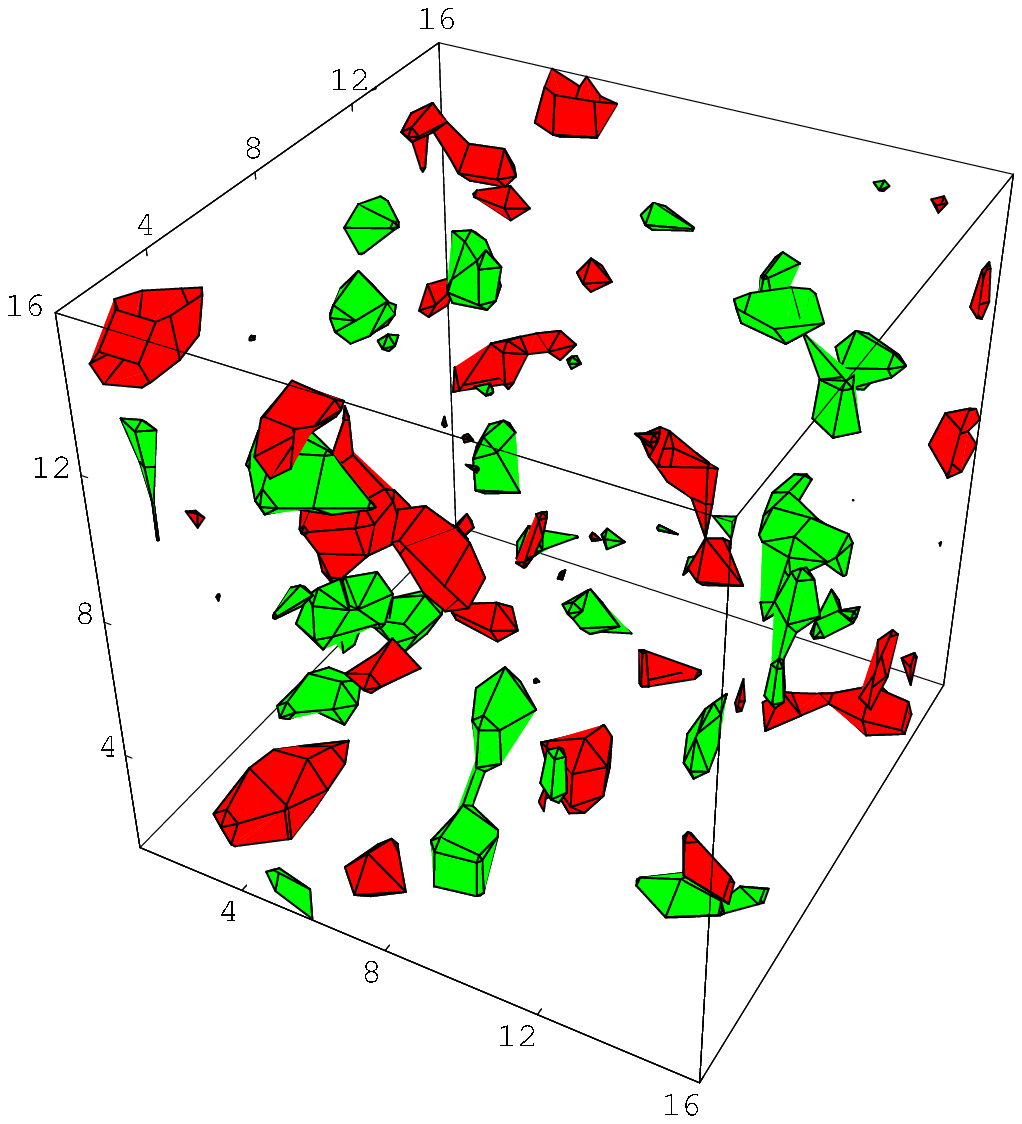}
    \hspace*{0.2cm}
    \includegraphics[scale=0.22]{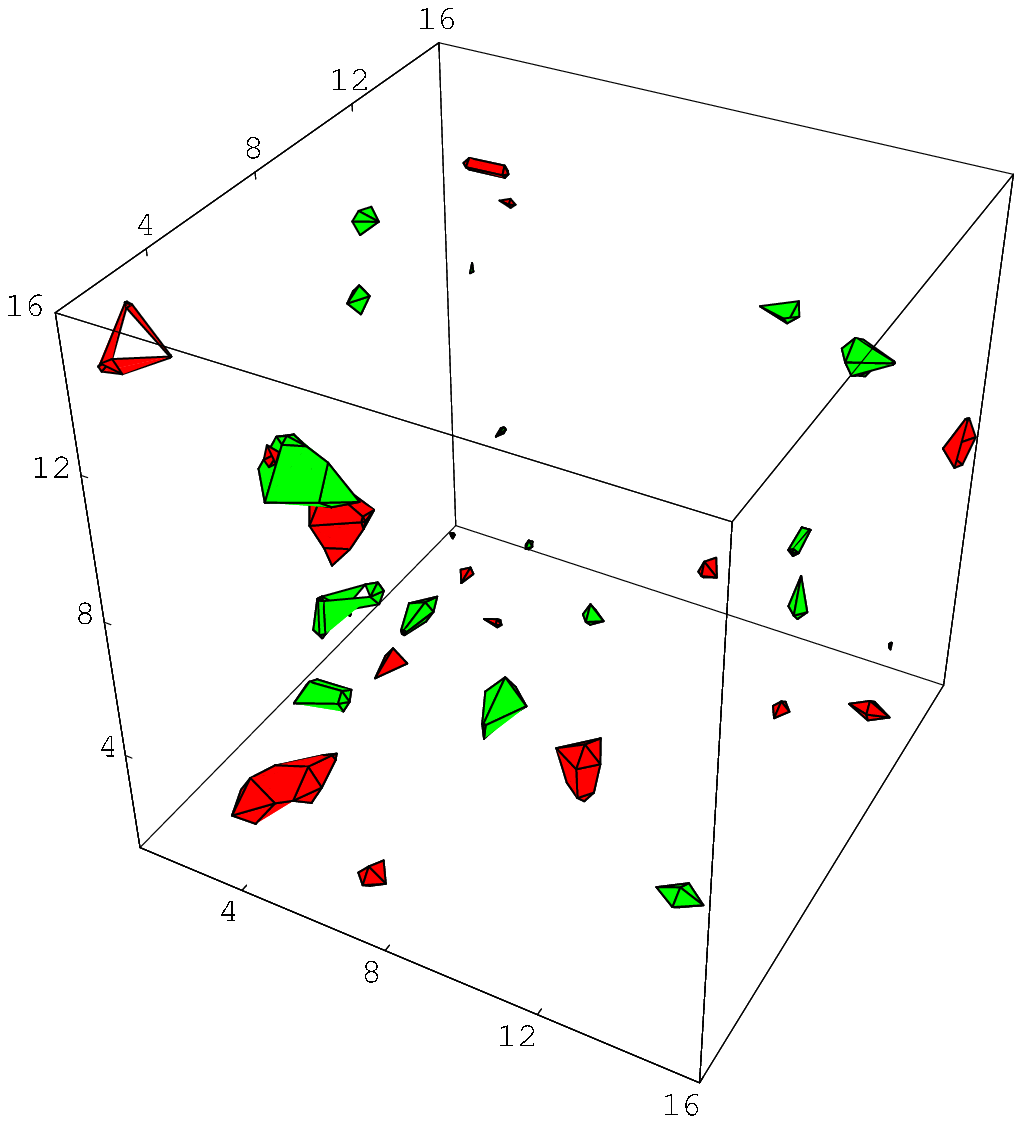}\\
    \includegraphics[scale=0.22]{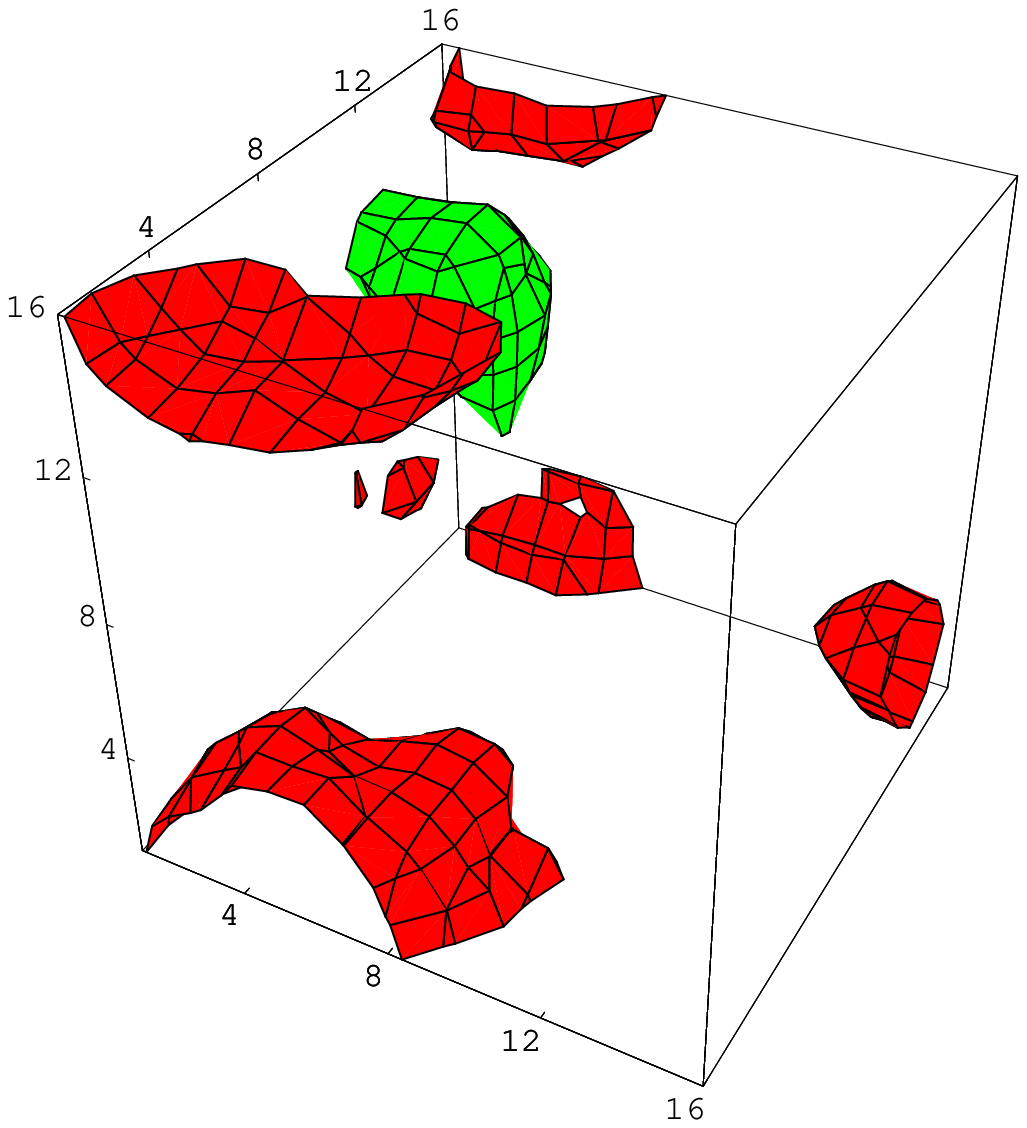}
    \hspace*{0.2cm}
    \includegraphics[scale=0.22]{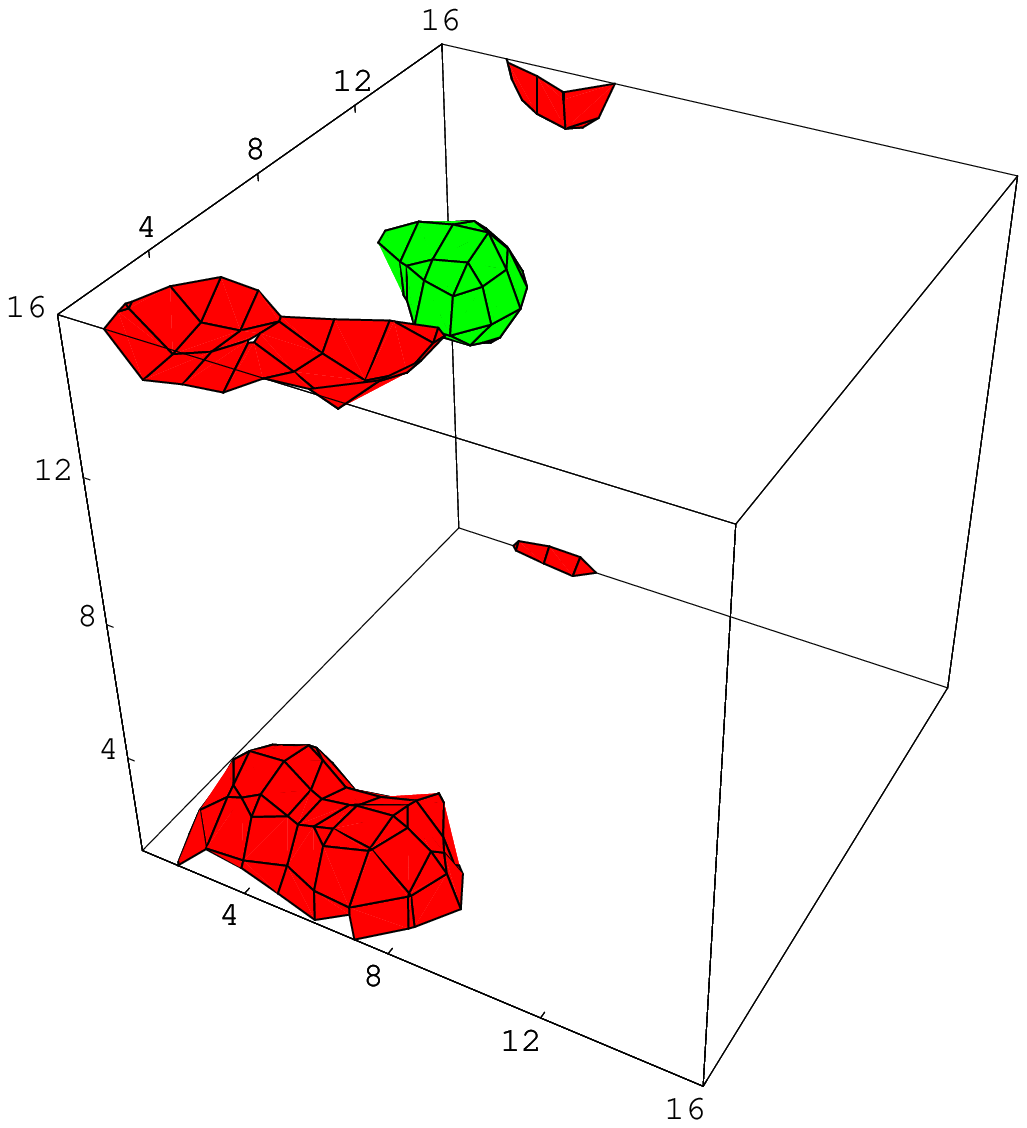}
    \hspace*{0.2cm}
    \includegraphics[scale=0.22]{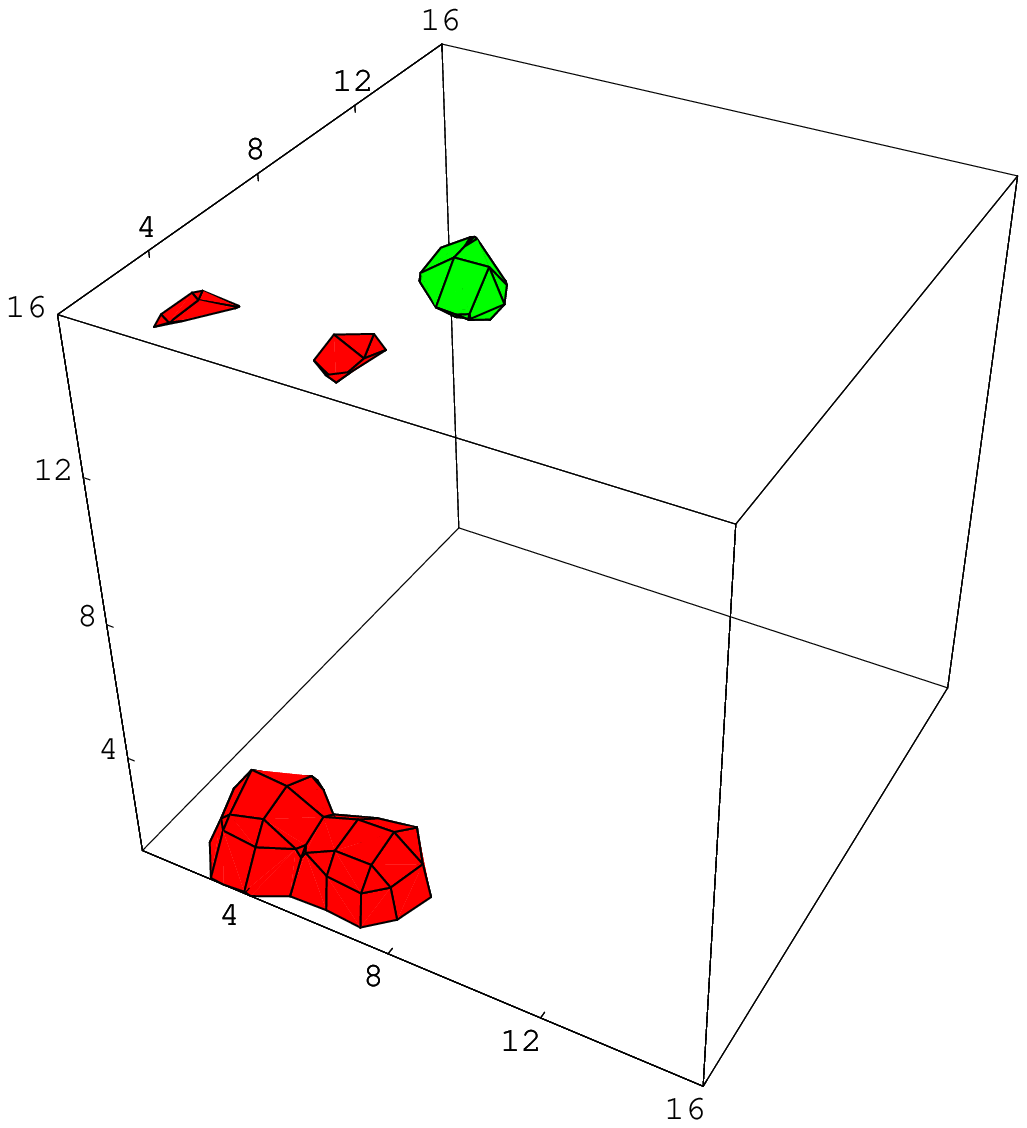}
    \hspace*{0.2cm}
    \includegraphics[scale=0.22]{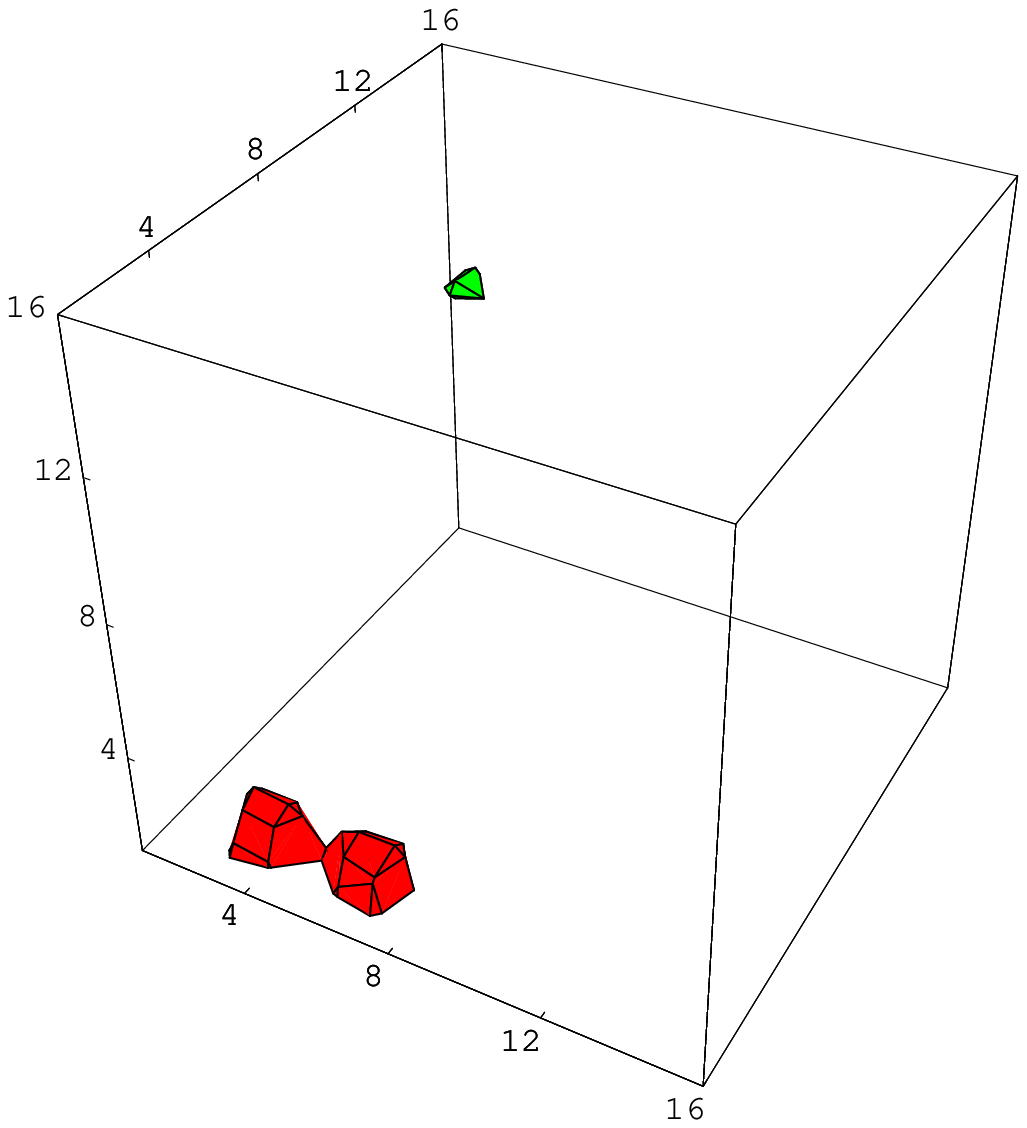}
\caption{Isosurfaces of topological charge density with 
$|q(x)|/q_{\rm max} = 0.1$, 0.2, 0.3, 0.4, in one timeslice of a 
single $16^3 \times 32$ configuration at $\beta=8.45$.
The upper pictures are based on the scale-$a$ density, 
the lower pictures on the eigenmode-truncated density with 
$a\, \lambda_{\rm sm} = 0.076$. Color encodes the sign of the 
charge enclosed by the isosurface. Fig. from~\cite{Ilgenfritz:2005hh}.}
\label{fig:Fig4}
\end{figure}

The meaning of the cluster analysis is geometrically visualized in 
Fig.~\ref{fig:Fig4}, again contrasting the scale-$a$ density with the filtered 
density. Both types of clusters are presented as a function of $q_{\rm cut}$. 
The upper row shows, say for $q_{\rm cut}/q_{\rm max} \approx 0.2$, that more and more 
irregular, spiky clusters of the unfiltered density are popping up before 
percolation sets in. The lower row demonstrates that the number of clusters 
of the filtered density remains small and practically independent of $q_{\rm cut}$. 
The density is ${\cal O}(1 {\rm~fm}^{-4})$, that means comparable with what 
has been estimated for the instanton density. Of course, with lowering $q_{\rm cut}$, 
the clusters slightly grow before they eventually merge. Coalescence happens 
finally below $q_{\rm cut}/q_{\rm max} = 0.1$. 
We see that the UV smoothed density exhibits cluster properties similar 
to the instanton model~\cite{Schafer:1996wv} or what cooling 
studies on the lattice have shown earlier~\cite{Negele:1998ev,GarciaPerez:2000hq}.
\vspace{-0.6cm}

\section{Fractal Dimensions}

\begin{figure}[b]
    \centering
    \includegraphics[scale=0.56,angle=0]{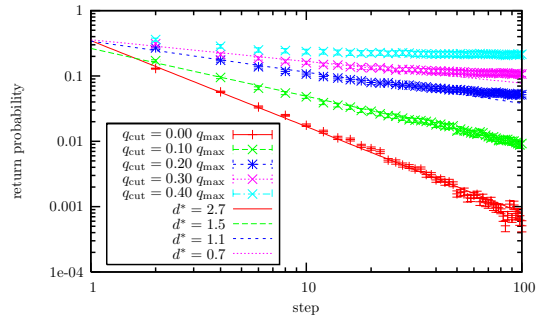}
\caption{The return probability of random walks in the biggest cluster of the 
scale-$a$ topological density as a function of the step number, and the fractal 
dimensions inferred from that for various lower cut-off values $q_{\rm cut}$.
Fig. from~\cite{Ilgenfritz:2007xu}.}
\label{fig:Fig5}
\end{figure}
The visualization in three dimensions does not immediately expose the (fractal) 
dimension of the clusters of the scale-$a$ topological density.
We have measured the (fractal) dimension of these clusters 
by a random walk method. While the cut-off 
$q_{\rm cut}$ defines the extension of the clusters as discussed above,  
a random walk is arranged inside the biggest cluster that is unable to penetrate 
the (fractal) border. A rough 
estimate of the dimensionality of the cluster at the respective 
level of $q_{\rm cut}$ 
can be inferred from the return probability to the starting point of the random 
walk. In $d^{*}$ dimensions it behaves as a function of the number of steps like
$P({\vec 0},\tau) = 1/(2\, \pi\, \tau)^{d^{*}/2}$. Figure~\ref{fig:Fig5} shows the 
observed power like behavior for different $q_{\rm cut}/q_{\rm max}$. 
The dimensions $d^{*}$ extracted from this study are shown in the legend. 
This analysis reveals a continuous change of the fractal dimension 
from $d^{*} < 1$ (characterizing the irregular spikes) to $d^{*} \approx 3$ 
(characterizing the $3D$ membranes) with a lowering cut-off 
$q_{\rm cut}/q_{\rm max}=0.4 \ldots 0.0$.
\vspace{-0.6cm}

\section{Smearing vs. Filtering}

\begin{figure}[ht]
    \centering
    \includegraphics[scale=0.24,angle=90]{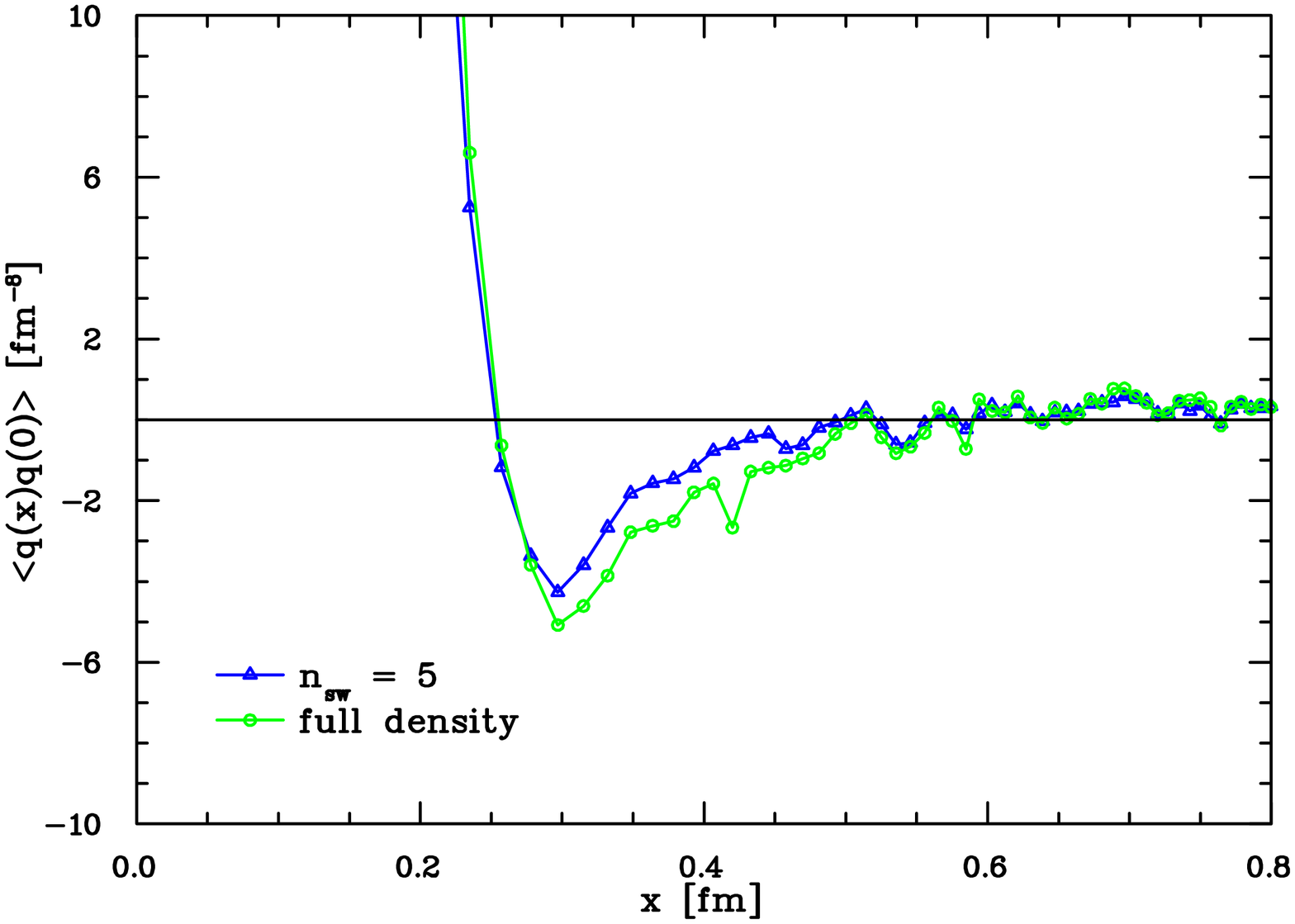}
    \hspace*{0.2cm}
    \includegraphics[scale=0.24,angle=90]{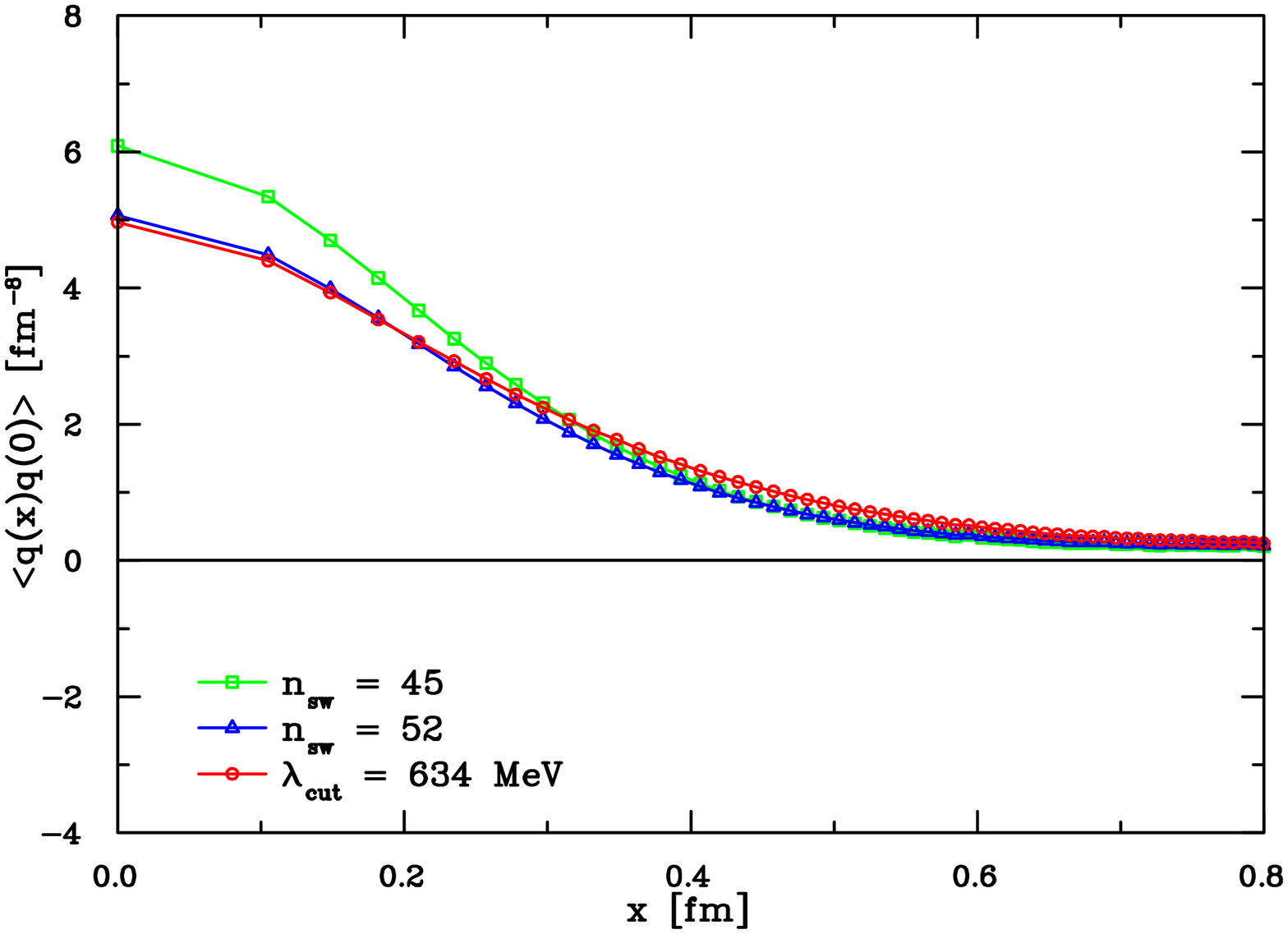} \\
    (a) ~~~~~~~~~~~~~~~~~~~~~~~~~~~~~~~~~~~~~~~ (b)
\caption{Left: the two-point function of the fermionic topological 
density at scale $a$ (green) compared with the bosonic definition 
after 5 steps of stout-link smearing (blue). 
Right: the two-point function of the fermionic topological density
with an UV cut-off at $\lambda_{\rm sm}=634 \mathrm{~MeV}$ (red) compared
with the bosonic definition after 52 steps of stout-link smearing (blue)  
when the correlator is fitted best. The best point-by-point matching 
of the density $q_{\lambda_{\rm sm}}$ with $\lambda_{\rm sm}=634 \mathrm{~MeV}$ 
is achieved with 45 steps. For this case of less smearing the correlator 
is steeper. Fig. from~\cite{Ilgenfritz:2008ia}.}
\label{fig:Fig6}
\end{figure}

\vspace{-0.2cm}
The similarity of the filtered topological density to clusters obtained
earlier by smearing methods can be studied more in detail by direct comparison. 
First let us recall that the overlap operator is {\it not ultralocal}. 
Even the scale-$a$ topological density given by overlap fermions represents some 
inherent non-locality of the overlap operator~\cite{Horvath:2005cv}. 
This is seen by 
considering the two-point function of the topological density. Theoretically, 
one expects that it is negative for all non-zero distances, with an infinite, 
$\delta$-like contact term at vanishing distances. The two-point function 
actually measured is shown as the green curve in Fig.~\ref{fig:Fig6}a. One needs 
to go to larger $\beta$ (a finer lattice) in order to see the positive core 
slowly shrinking and the minimum becoming deeper.
To compare with smearing, the alternative measurement of the topological density 
(and of the two-point function) uses an improved bosonic definition of the 
topological charge density, essentially
\begin{equation}
q(x) = \frac{1}{16\, \pi^2} {\rm tr} 
\left( {\vec E(x)} \cdot {\vec B(x)} \right) \, ,
\label{eq:EtimesB}
\end{equation}
in terms of an highly improved electric and 
magnetic field strength~\cite{BilsonThompson:2002jk},
that includes plaquettes and bigger loops. To apply it to
a generic lattice configuration requires a few iterations of smearing (actually, 
of stout link smearing with respect to an overimproved action~\cite{Moran:2008ra})
before integer-valued topological charges $Q$ are measured with good precision. 
Figure 6a shows that the overlap definition and the bosonic 
definition of the topological density agree best to reproduce a similar two-point 
function at 5 smearing steps. Figure 6b 
shows the gluonic two-point function after 45 and 52 smearing steps compared 
with the overlap definition of the two-point function with $\lambda_{\rm sm}=634$ MeV. 
At this level of smearing, the negativity is already washed out and extended 
clusters are visible. A point-by-point matching of the topological density and a 
matching of the two-point function is achieved with almost the same number of 
smearing iterations.

\begin{figure}[ht]
    \centering
    \includegraphics[scale=0.24,angle=90]{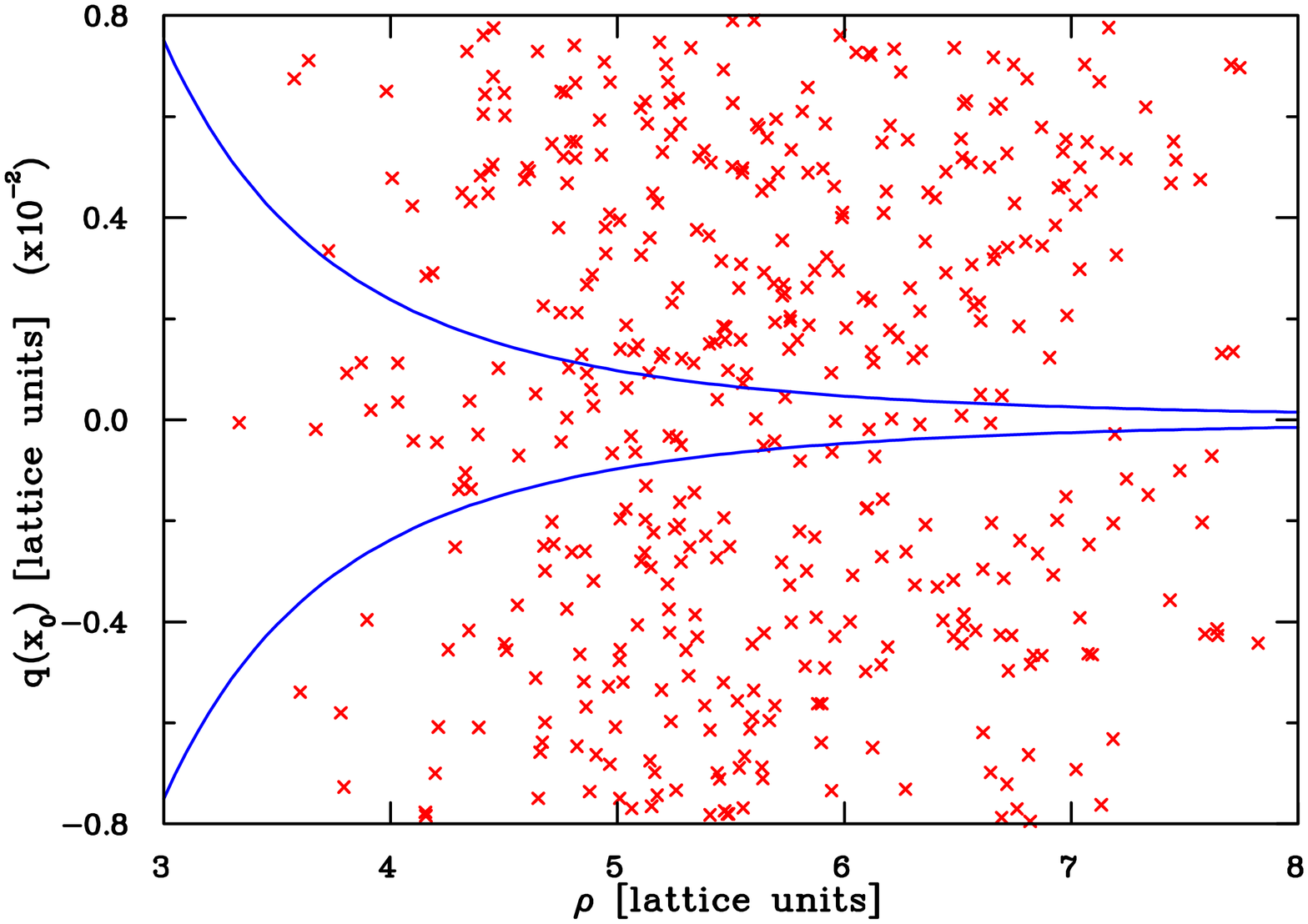}
    \hspace*{0.2cm}
    \includegraphics[scale=0.24,angle=90]{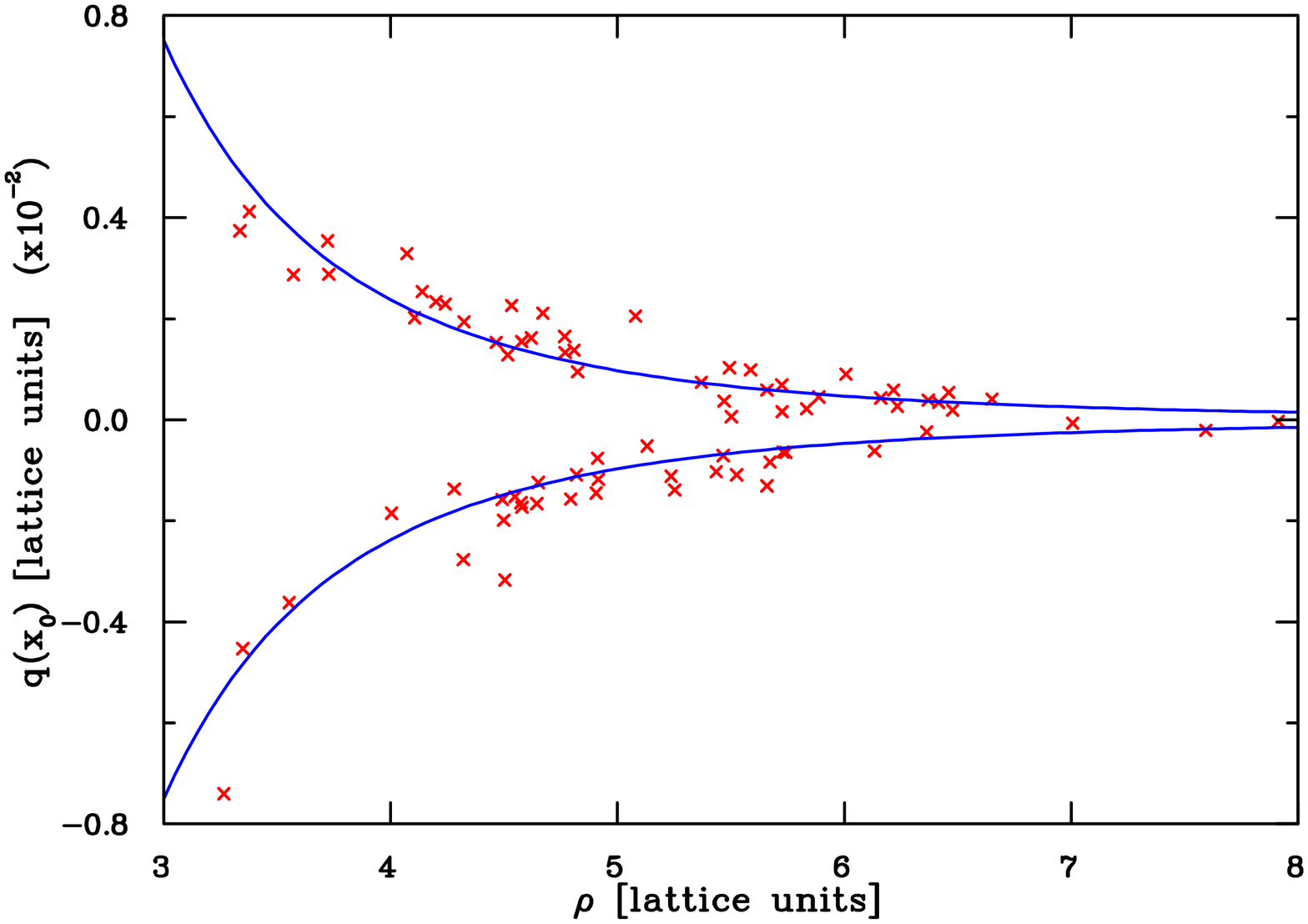} \\
    (a) ~~~~~~~~~~~~~~~~~~~~~~~~~~~~~~~~~~~~~~~ (b)
\caption{The cluster content after 5 and 40 smearing steps in the 
$q_{\rm max}$-$\rho_{\rm inst}$ plane. Fig. from~\cite{Ilgenfritz:2008ia}.}
\label{fig:Fig7}
\end{figure}

\begin{figure}[h]
    \centering
    \includegraphics[scale=0.27,angle=0]{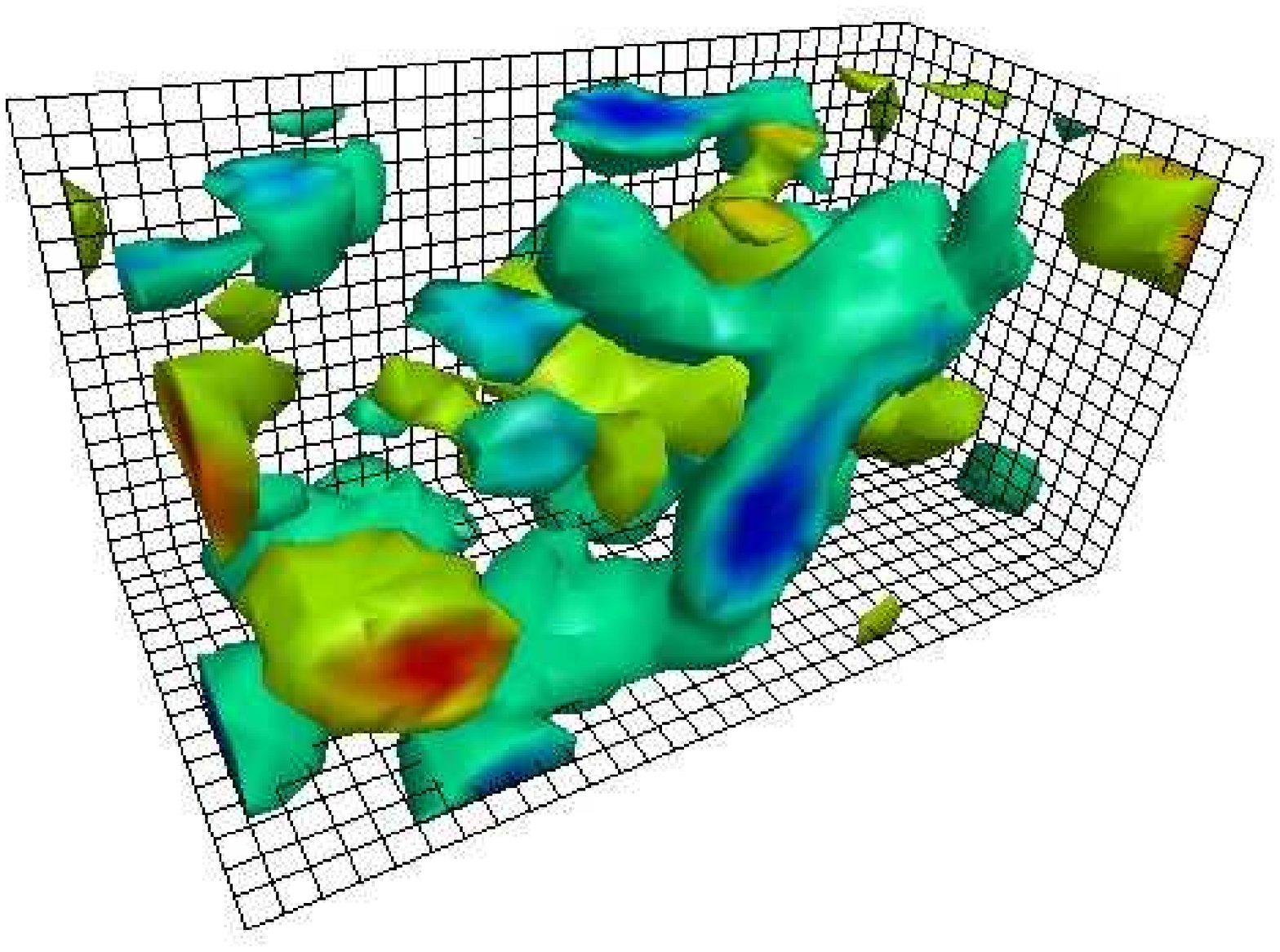}
    \hspace*{0.2cm}
    \includegraphics[scale=0.27,angle=0]{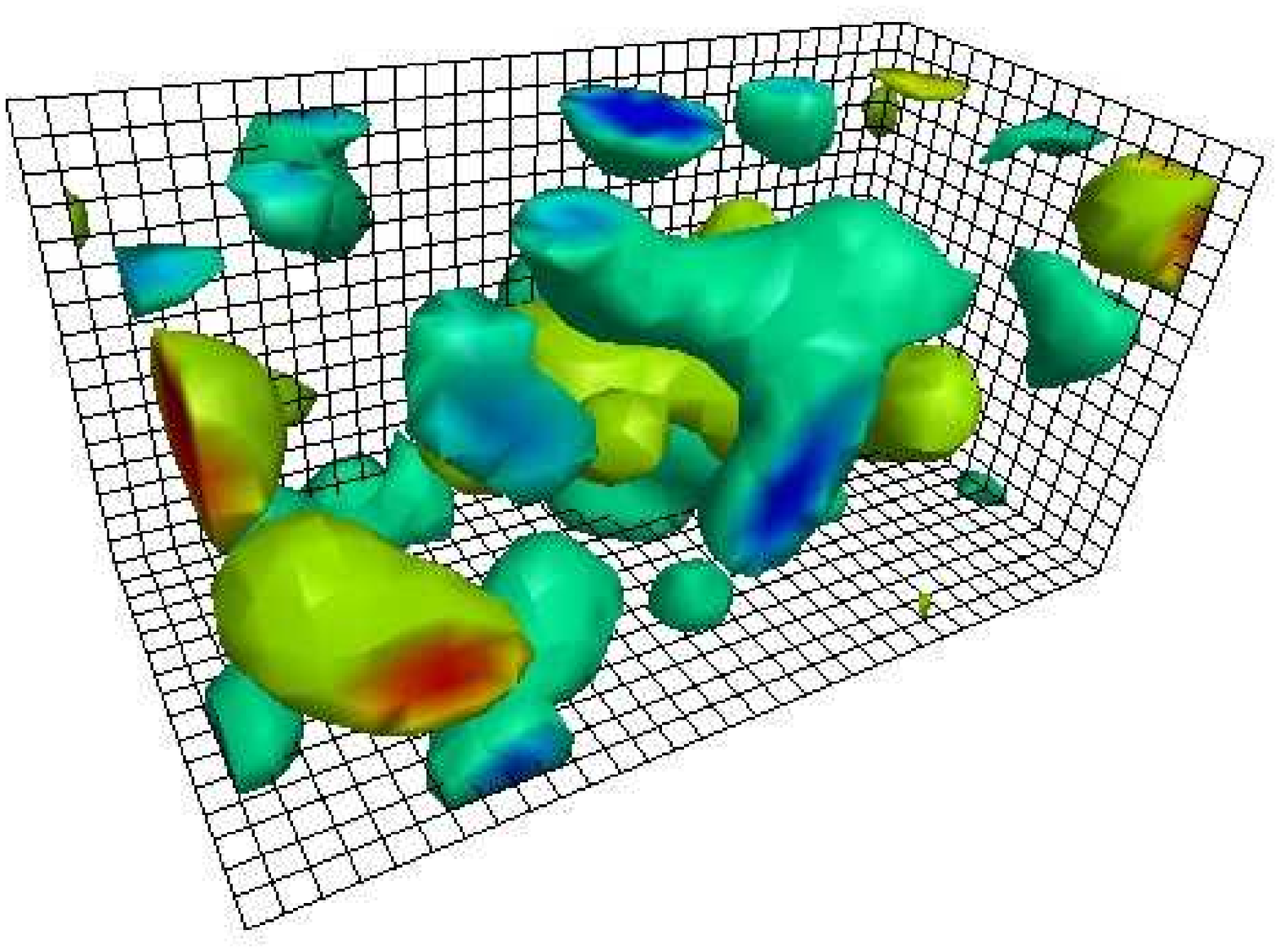} \\
    (a) ~~~~~~~~~~~~~~~~~~~~~~~~~~~~~~~~~~~~~~~ (b)
\caption{The fermionic topological charge density of a $Q = 0$
configuration with $\lambda_{\rm sm}=634 \mathrm{~MeV}$ (left)
compared with the bosonic one after 48 sweeps of stout-link 
smearing (right). Negative: blue/green, positive: red/yellow. 
Fig. from~\cite{Ilgenfritz:2008ia}.}
\label{fig:Fig8}
\end{figure}

There are two parameters, $q_{\rm max}$ and the size $\rho_{\rm inst}$ 
(historically the 
instanton radius) suitable to characterize each cluster. Now the size parameter 
$\rho_{\rm inst}$ is estimated by the curvature of $q(x)$ close to the maxima $x_0$ 
where $q(x_0)=q_{\rm max}$. In the form of a scatter plot in the plane spanned by these
two parameters one can describe and compare the cluster composition of the
gauge field.
In Fig.~\ref{fig:Fig7} this is shown after 5 and 40 stout-link smearing steps. 
After 40 smearing steps the cluster multiplicity is essentially reduced, while 
an instanton-like relation (the curves drawn in Fig.~\ref{fig:Fig7})
\begin{equation}
q_{\rm max} = \frac{6}{\pi^2\, \rho_{\rm inst}^4}  \, ,
\label{eq:instantonlike}
\end{equation}
is enforced within some tolerance. At the same time the clusters are far from 
being ideal O(4) symmetric instanton solutions. 
The bosonic and the fermionic view of 
the cluster structure is shown in the left and right panel of Fig.~\ref{fig:Fig8}.
They show a remarkable similarity.
\vspace{-0.6cm}

\section{Selfduality}

\begin{figure}[b]
    \centering
\vspace{-0.6cm}
    \includegraphics[scale=0.28,angle=-90]{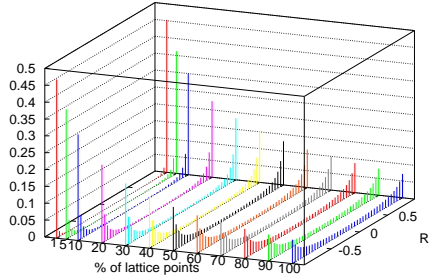}
\caption{Normalized histogram with respect to the UV smoothed (anti)selfduality $R$ 
for different subsamples of lattice points ($1 \%$, $5 \%$, $10 \%$ etc.) ordered 
with respect to the UV smoothed action. $20$ overlap modes have been included in the
smoothing. Fig. from~\cite{Ilgenfritz:2007xu}.}
\label{fig:Fig9}
\end{figure}
This leads to the question to what extent the UV smoothed {\it fieldstrength tensor}
is {\it locally} selfdual or antiselfdual, or even a semiclassical (instanton) 
solution.
Gattringer was the first to ask this question in Ref.~\cite{Gattringer:2002gn}.
We have applied his technique using the eigenmodes of the overlap Dirac operator.
Truncating the sum over eigenvalues, an UV filtered form of the field strength
$F_{\mu\nu}$ can be obtained that allows to evaluate the respective infrared
(IR), UV smoothed topological charge density,
\begin{equation}
q_{IR}(x) \propto {\rm Tr} \left( F_{\mu\nu}(x)\, \tilde{F}_{\mu\nu}(x) \right)  \, ,
\label{eq:QIR}
\end{equation}
and action density
\begin{equation}
s_{IR}(x) \propto {\rm Tr} \left( F_{\mu\nu}(x)\, F_{\mu\nu}(x) \right) \, .
\label{eq:SIR}
\end{equation}
Analogously to the local chirality of the non-zero modes one can define
the ratio
\begin{equation}
r(x) = \frac{s_{IR}(x) - q_{IR}(x)}{s_{IR}(x) + q_{IR}(x)}  \, ,
\label{eq:RATIO}
\end{equation}
which can be converted to
\begin{equation}
R(x) = \frac{4}{\pi} \arctan \left( \sqrt{r(x)} \right) - 1 \in [-1,+1] \, .
\label{eq:R}
\end{equation}
Regions with $R(x) \approx -1$ or $R(x) \approx +1$ are characterized by 
the field strength being locally selfdual (antiselfdual). 
Figure~\ref{fig:Fig9} shows histograms 
with respect to $R$ for the whole lattice and for subsets of lattice points 
selected by ranking according to the action density. With higher action
density (say, for $ < 30 \%$ of the lattice points forming ``hot spots''
of filtered action $S_{IR}$) the parameter $R$ is
distributed close to $\pm 1$ (i.e. $|R| > 0.7$). 
Clusters with respect to $|R| > 0.98$ are overlapping with clusters of a 
suitably filtered topological density.
\vspace{-0.6cm}

\section{Localization and Local Chirality of Overlap Eigenmodes}

\vspace{-0.2cm}
\begin{figure}[h]
    \centering
    \includegraphics[scale=0.55]{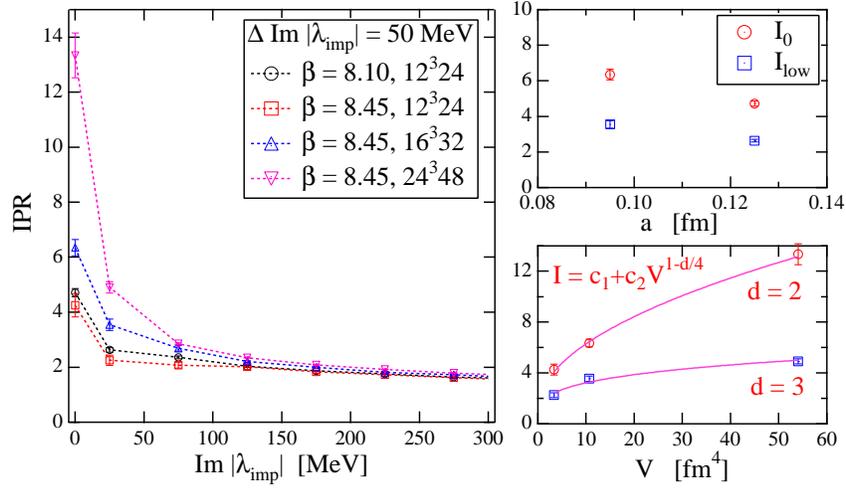}
\caption{IPR: the dependence on $\lambda$ (left),
the $a$-dependence (right upper), and the $V$-dependence (right lower).
Fig. from~\cite{Koma:2005sw}.}
\label{fig:Fig10}
\end{figure}
We have also analyzed the localization behavior of individual overlap eigenmodes. 
The interest in the localization properties of Dirac and scalar eigenmodes has been
discussed by de Forcrand~\cite{deForcrand:2006my}.
In the left panel of Fig.~\ref{fig:Fig10} we show the average inverse 
participation ratio (IPR) 
typical for zero modes and for bins of eigenvalues in dependence on coarseness 
and lattice size. The localization grows approaching the continuum limit and 
with increasing physical volume. The right panel shows, separately for zero 
modes and the first bin ($|\lambda| <$  50 MeV), the change of the IPR with 
$a$ and with $V$. From fits we have concluded that zero modes are typically 
2-dimensionally extended, while the lowest non-zero modes are 3-dimensional 
objects.

\begin{figure}[h]
    \centering
    \includegraphics[scale=0.8]{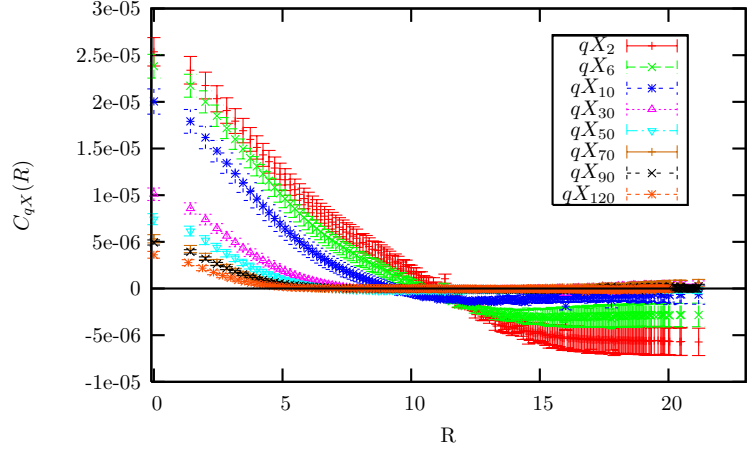}
\caption{The correlation function between the local chirality $X(x)$
of selected eigenmodes and the filtered topological density $q_{\lambda}(y)$
in a configuration with topological charge $Q=0$. 
Fig. from~\cite{Ilgenfritz:2007xu}.}
\label{fig:Fig11}
\end{figure}
Similar to (\ref{eq:RATIO}) the local chirality $X(x)$ of the non-zero modes can
be defined. This quantity is highly correlated with the UV smoothed topological 
density. For $\lambda_{\rm sm}=200 {\rm~MeV}$ the correlation function between
$X(x)$ and $q_{\lambda_{\rm sm}}(y)$ is shown in Fig.~\ref{fig:Fig11}. 
The correlator is positive for the lowest 120 eigenmodes within a range of 
distance $R < {\cal O}(1 {\rm~fm})$.
\vspace{-0.8cm}

\section{Technical Details}

To compute the low lying eigenvalues of the overlap Dirac operator
(\ref{eq:OverlapDirac}) we use an adaption of the implicitly restarted Arnoldi method which is used in the
ARPACK \cite{ARPACK} library. The advantage of this method is that 
the action of a matrix on a vector can be computed freely without the need to
express the input matrix explicitly.

The main computational challenge using the overlap Dirac operator as an input
matrix is the computation of the sign function ${\rm sgn}(D_W)$ in
(\ref{eq:OverlapDirac}). We project out the lowest ${\cal O}$(50) eigenvalues
of $D_W$ and treat them exactly. The rest acting in the orthogonal subspace is
approximated using minmax polynomials~\cite{Giusti:2002sm}.
The application of the Wilson-Dirac operator $D_W$  on a vector, the main
computational kernel of almost every lattice QCD code,  is implemented 
in assembler and highly optimized using SSE2 instructions. The data is 
aligned on full cache lines and stored in a special way to make optimal use of
the 128-bit XMM registers. Prefetch instructions are issued to move the data to the
cache accurately timed. The SSE2 version reaches approx. $1/3$ of peak
performance on a $16^3 \times 32$ lattice on the Linux cluster of the LRZ. 

The computation of the scale-$a$ topological density $q(x)$
requires to evaluate the local trace over Dirac and color indices of the
overlap Dirac operator  with the $\gamma_5$ matrix included according to (\ref{eq:TopDensI}). 
This can be trivially parallelized, but is numerically extremely expensive,
since the overlap operator has to be applied on $12\; V_{\rm lat}$ unit vectors.
To compute $q(x)$ on one single configuration  on a $16^3 \times 32$ lattice
\linebreak approx. $1.3\cdot10^4$ CPU hours for
totally 1.6 million applications of the overlap Dirac operator were required.

\section{Conclusions}

We have confirmed that the scale-$a$ topological density is highly singular and
the topological charge of either sign is globally filling the two 
extended $3d$ percolating structures. A similar low-dimensionality is found 
for chiral zero-modes (two-dimensional) and the lowest non-chiral non-zero 
modes (three-dimensional). The microscopic topological density is correlated 
with vortices and monopoles~\cite{Ilgenfritz:2007ua}.

The UV filtered topological density shows cluster properties reminiscent of
instantons. We found a highly correlated behavior of a huge number of lowest
modes resembling the presence of a semiclassical background. The suitably UV 
smoothed field strength tensor becomes selfdual (antiselfdual) at ``hot spots'' 
where the corresponding action becomes maximal. The reason for this 
self-organization between the fermionic modes is still unclear.

\bibliographystyle{spmpsci}

\end{document}